\documentclass[12pt,preprint]{aastex}

\begin{document}
\def\ale{\mathrel{\hbox{\rlap{\hbox{\lower4pt\hbox{$\sim$}}}\hbox{$<$}}}}
\def\age{\mathrel{\hbox{\rlap{\hbox{\lower4pt\hbox{$\sim$}}}\hbox{$>$}}}}

\title{GRB~011121: A Massive Star Progenitor}

\author{
P.~A.~Price\altaffilmark{1,2},
E.~Berger\altaffilmark{2},
D.~E.~Reichart\altaffilmark{2},
S.~R.~Kulkarni\altaffilmark{2}, 
R.~Subrahmanyan\altaffilmark{3},
R.~M.~Wark\altaffilmark{3},
M.~H.~Wieringa\altaffilmark{3},
D.~A.~Frail\altaffilmark{4,2},
J.~Bailey\altaffilmark{5},
B.~Boyle\altaffilmark{5},
E.~Corbett\altaffilmark{5},
K.~Gunn\altaffilmark{6},
S.~D.~Ryder\altaffilmark{5},
N.~Seymour\altaffilmark{6},
K.~Koviak\altaffilmark{7},
P.~McCarthy\altaffilmark{7},
M.~Phillips\altaffilmark{7},
T.~S.~Axelrod\altaffilmark{1},
J.~S.~Bloom\altaffilmark{2},
S.~G.~Djorgovski\altaffilmark{2},
D.~W.~Fox\altaffilmark{2},
T.~J.~Galama\altaffilmark{2},
F.~A.~Harrison\altaffilmark{2},
K.~Hurley\altaffilmark{8},
R.~Sari\altaffilmark{2},
B.~P.~Schmidt\altaffilmark{1},
S.~A.~Yost\altaffilmark{2},
M.~J.~I.~Brown\altaffilmark{9},
T.~Cline\altaffilmark{10},
F.~Frontera\altaffilmark{11},
C.~Guidorzi\altaffilmark{12} and
E.~Montanari\altaffilmark{12}.
}

\altaffiltext{1}{Research School of Astronomy \& Astrophysics, Mount Stromlo
Observatory, via Cotter Road, Weston, ACT, 2611, Australia.}
\altaffiltext{2}{Palomar Observatory, 105-24, California Institute of
Technology, Pasadena, CA, 91125.}
\altaffiltext{3}{Australia Telescope National Facility, CSIRO, P.O. Box 76,
Epping NSW 1710, Australia.}
\altaffiltext{4}{National Radio Astronomy Observatory, P.O. Box O, Socorro,
NM, 87801.}
\altaffiltext{5}{Anglo-Australian Observatory, P.O. Box 296, Epping,
NSW 1710, Australia.}
\altaffiltext{6}{Department of Physics and Astronomy, University of
Southampton, Highfield, Southampton SO17 1BJ, United Kingdom.}
\altaffiltext{7}{Carnegie Observatories, 813 Santa Barbara Street, Pasadena,
CA 91101.}
\altaffiltext{8}{University of California Space Sciences Laboratory,
Berkeley, CA, 94720.}
\altaffiltext{9}{National Optical Astronomy Observatory, P.O. Box 26732,
Tucson, AZ, 85726.}
\altaffiltext{10}{NASA Goddard Space Flight Center, Code 661, Greenbelt,
MD 20771.}
\altaffiltext{11}{Istituto Tecnologie e Studio Radiazioni Extraterrestri,
CNR, Via Gobetti 101, 40129 Bologna, Italy.}
\altaffiltext{12}{Dipartimento di Fisica, Universita di Ferrara, Via
Paradiso 12, 44100, Ferrara, Italy.}

\begin{abstract}
Of the cosmological gamma-ray bursts, GRB~011121 has the lowest
redshift, $z=0.36$.  More importantly, the multi-color excess in the
afterglow detected in the Hubble Space Telescope (HST) light curves is
compelling observational evidence for an underlying supernova. Here we
present near-infrared and radio observations of the afterglow. We
undertake a comprehensive modeling of these observations and those
reported in the literature and find good evidence favoring a wind-fed
circumburst medium. In detail, we infer the progenitor had a mass loss
rate of $\dot M\sim 10^{-7}/v_{w3}\,M_\odot\,{\rm yr}^{-1}$ where
$v_{w3}$ is the speed of the wind from the progenitor in units of
$10^3$ km s$^{-1}$.  This mass loss rate is similar to that inferred
for the progenitor of~SN 1998bw which has been associated with
GRB~980425.  Our data, taken in conjunction with the HST results of
\citet{bkp+02}, provide a consistent picture: the long duration GRB
011121 had a massive star progenitor which exploded as a supernova at
about the same time as the GRB event.

\end{abstract}



\section{Introduction}
\label{sec:introduction}

On 2001 November 21 at 18:47:21 UT, GRB 011121 was detected and
localized by the Italian-Dutch satellite BeppoSAX \citep{piro01a}.
The localization was further improved by the InterPlanetary Network
\citep{hurley+01} and an optical transient was identified by the OGLE
group \citep{wsg01}.  Spectroscopy of the transient revealed emission
lines interpreted as arising from the host galaxy at a redshift of
$z=0.36$ \citep{igsw01}.

Low redshift GRBs are particularly valuable in uncovering the origin
of GRBs. If GRBs result from the death of massive stars then it is
reasonable to expect an underlying supernova (SN).  \citet{bkd+99}
attributed a late-time red excess seen in the afterglow emission of
GRB~980326 to an underlying SN. This result triggered searches for
similar excesses with no clear success save GRB 970228
\citep{rei99,gtv+00}.  The low redshift is critical to such searches
since the SN contribution is expected to exhibit strong absorption
below 4000 \AA\ (see \citealt{bkd+99}).

Given this motivation, we triggered a sequence of multi-color and
multi-epoch Wide Field and Planetary Camera 2 (WFPC-2) observations
with the {\it Hubble Space Telescope} (HST).  \citet{ghj+02} noted
that the $R$-band flux of the first epoch of the HST observations was
significantly in excess of the extrapolation of the power law decay of
the early ground-based optical afterglow and attributed this to an
underlying SN component. In Paper I (\citealt{bkp+02}) we presented
four-epoch multi-color HST light curves and show the data are
explained by an underlying SN similar to SN 1998bw \citep{gvv+98}
except fainter by about 2/3 magnitude. At this point, there appears to
be compelling evidence for GRB~011121 to be associated with a SN which
exploded at about the same time as the gamma-ray event \citep{bkp+02}.

This GRB-SN link is an essential expectation in the collapsar model
\citep{woo93} in which GRBs result from the death of certain massive
stars. Another essential consequence of any massive star origin for
GRBs, as noted by \citet{cl99}, is a circumburst medium fed by the
inevitable and copious mass loss suffered by massive stars throughout
their lives.  Afterglow observations are excellently suited to
determining not only the geometry of the explosion but also the
distribution of circumburst matter.  Unfortunately, until now there
has been no clear evidence for a wind-fed circumburst medium (density,
$\rho\propto r^{-s}$ with $s\sim 2$; here $r$ is the distance from the
explosion site) in the afterglow of any cosmologically located GRB.

Here we report near-infrared (NIR) and radio observations of the
afterglow of GRB\,011121.  We undertake afterglow modeling of this
important event and to our delight have found a good case for a
wind-fed circumburst medium. Thus, the totality of the data --- the HST
optical lightcurves and multi-wavelength (radio, NIR, and optical)
data --- now support a massive star origin for this GRB.

\section{Observations}
\label{sec:obs}

\noindent{\bf Gamma-Rays:} GRB~011121 was observed by numerous
spacecraft in the InterPlanetary Network: Ulysses, BeppoSAX (GRBM),
HETE-2 (FREGATE), Mars Odyssey (HEND) and Konus-Wind.
The $T_{90}$ duration, as determined from the Ulysses data, was 28 s,
placing this event in the class of "long bursts"
(Figure~\ref{fig:grb-lc}).  The peak flux in the 25--100 keV range,
over 0.25 s, was $2.4 \times 10^{-6}$ erg\,cm$^{-2}$\,s$^{-1}$, and
the fluence was $2.4 \times 10^{-5}$ erg\,cm$^2$.


\noindent{\bf Near-Infrared:} We observed the afterglow in the
near-infrared with the newly-commissioned IRIS2 on the
Anglo-Australian Telescope (AAT), WFIRC on the du~Pont 2.5-m telescope
and the IRCam on the Walter Baade 6.5-m telescope in $J$ and $K_s$
filters.  The images were dark-subtracted, flat-fielded,
sky-subtracted and combined using DIMSUM \citep{edsw+99} in IRAF.
PSF-fitting photometry of the afterglow using DAOPHOT
\citep{stetson87} was performed relative to point sources in the
field.  Our multiple calibrations are consistent with each other and
we estimate the systematic error to be less than 0.05 mag (see
Table.~\ref{tab:nir}).

\noindent{\bf Radio:} We initiated observations of GRB~011121 with the
Australia Telescope Compact Array (ATCA) starting on 2001 November
22.58 UT (see Table~\ref{tab:radio}).  The data were reduced and
imaged using the Multichannel Image Reconstruction, Image Analysis and
Display (MIRIAD) software package.

\section{Modeling the Afterglow}
\label{sec:model}

\subsection{Dust Extinction}
\begin{figure}[tbp]
\plotone{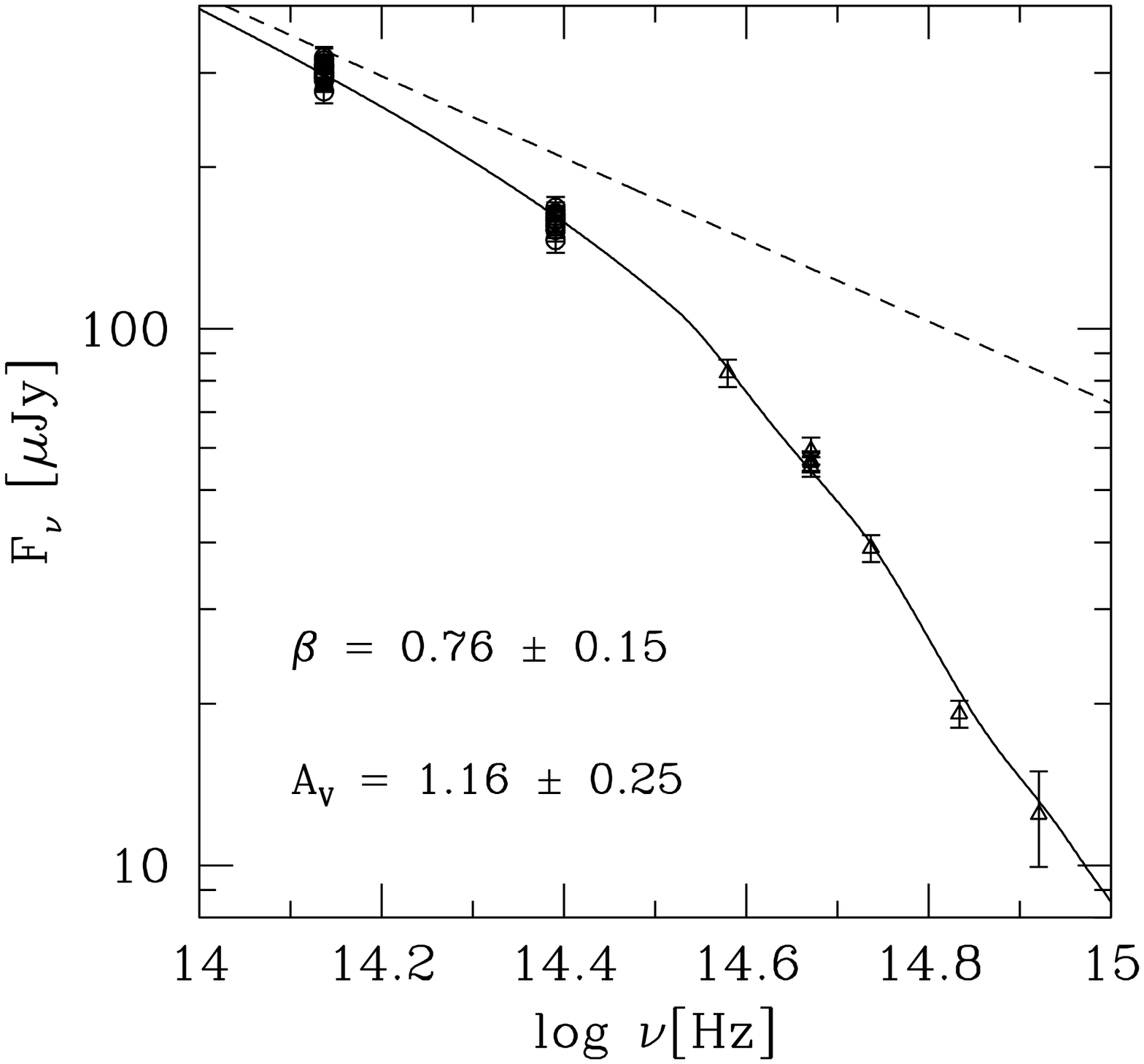}
\caption{The optical and NIR spectral flux distribution of the
afterglow of GRB~011121 at 0.5 d, based on measurements presented here
(circles) and from the literature (triangles; \citealt{obs+01,sw01}).
Measurements taken within $0.5\pm\ 0.1$ d have been transformed to 0.5
d using the best fit model.  The solid lines indicates our best fit to
the data, using a power-law model plus foreground extinction.  The
dashed line is the intrinsic spectrum of the afterglow.}
\label{fig:sfd}
\end{figure}

In Figure \ref{fig:sfd} we display the optical/NIR spectrum of
GRB~011121.  The apparent curvature in the spectrum indicates a large
magnitude of dust extinction.  In view of this, estimating the dust
extinction accurately is critical not only for the afterglow modeling
but also as an important input parameter for the supernova modeling of
the HST lightcurves \citep{bkp+02}.

From the IR dust maps \citep{sfd98} we estimate $A_V \approx 1.6$
mag. However, the IR maps have low angular resolution. Indeed, it
appears that the line-of-sight to the afterglow passes through the
edge of a dust cloud $\sim 45$ arcmin in extent.  Fortunately, the
availability of both the optical and NIR afterglow data allow us to
directly estimate the extinction along this line of sight directly.

We make the reasonable assumption that the optical/NIR afterglow
follows the standard power-law model, $F_\nu \propto t^{-\alpha}
\nu^{-\beta}$, and we apply the parametric extinction curves of
\citet{ccm89} and \citet{fm88} along with the interpolation suggested
by \citet{rei01a}.  Thanks to the abundance of our NIR data, which
suffers little extinction, we can break the degeneracy between $\beta$
and the magnitude of the extinction, $A_{\rm V}$.


In addition to our own measurements we have included those reported in
the literature (and noted in Figure~\ref{fig:sfd}).  Since late-time
measurements are increasingly dominated by an uncertain mix of the
afterglow, the host galaxy and the nearby star B \citep{bkp+02} we
restrict the analysis to data obtained over the first two days.

Our best fit has an unacceptable $\chi^2$ = 66 for 48 degrees of
freedom, but this is mainly dominated by outliers, particularly in the
data from the AAT where the seeing blended star B with the afterglow
in some observations.  Inserting an additional 3\% error decreases the
$\chi^2$ to match the number of degrees of freedom. The additional
error term, while {\it ad hoc}, is reasonable given the variety of
telescopes and reduction techniques in our data set.

Our measured extinction is $A_V = 1.16 \pm 0.25$~mag, distinctly
lower than that deduced from the dust maps.  The type of extinction
curve (e.g., Milky Way, LMC, SMC etc.) is unconstrained by these
observations.  We have not solved for extinction within the host
galaxy, but the off-center location of the GRB \citep{bkp+02} makes it
likely that the contribution from extinction within the host galaxy is
small.  Finally, we measure $\alpha = 1.66 \pm 0.06$ and $\beta = 0.76
\pm 0.15$, without assuming any specific afterglow model.

\subsection{Afterlow Models}

Armed with $\alpha$ and $\beta$ we now consider three afterglow
models: (i) isotropic expansion into a homogeneous medium
\citep{spn98}, (ii) isotropic expansion into a wind-stratified medium
\citep{cl99}, and (iii) collimated expansion into a homogeneous or
wind-stratified medium \citep{sph99}.  The models can be distinguished
by a closure relation, $\alpha+b\beta+c=0$.  These closure relations
are due to the dependence of both $\alpha$ and $\beta$ on the electron
energy distribution index, $p$, and the values of $b$ and $c$ depend
on the location of the cooling frequency, $\nu_c$, relative to the
optical/NIR frequency, $\nu_O$, at the epoch of the observations.

As can be seen from Table~\ref{tab:p}, models with isotropic expansion
into an homogeneous medium (or, equivalently, a jet which becomes
apparent on a timescale longer than the epochs of the optical/NIR data
used here, $t_j\age 2$ d) are ruled out by the closure relations at
more than $2\sigma$ significance.  Two models produce closure
consistent with zero: ({\it A}) A wind model with $\nu_c>\nu_O$
(effective epoch day 1), and $p=2.55\pm 0.08$; and ({\it B}) A fully
developed jet at the time of the first optical observation, $t_j<0.5$
d, with $\nu_c<\nu_O$ and $p=1.66\pm 0.06$.

\begin{figure}[tbp]
\plotone{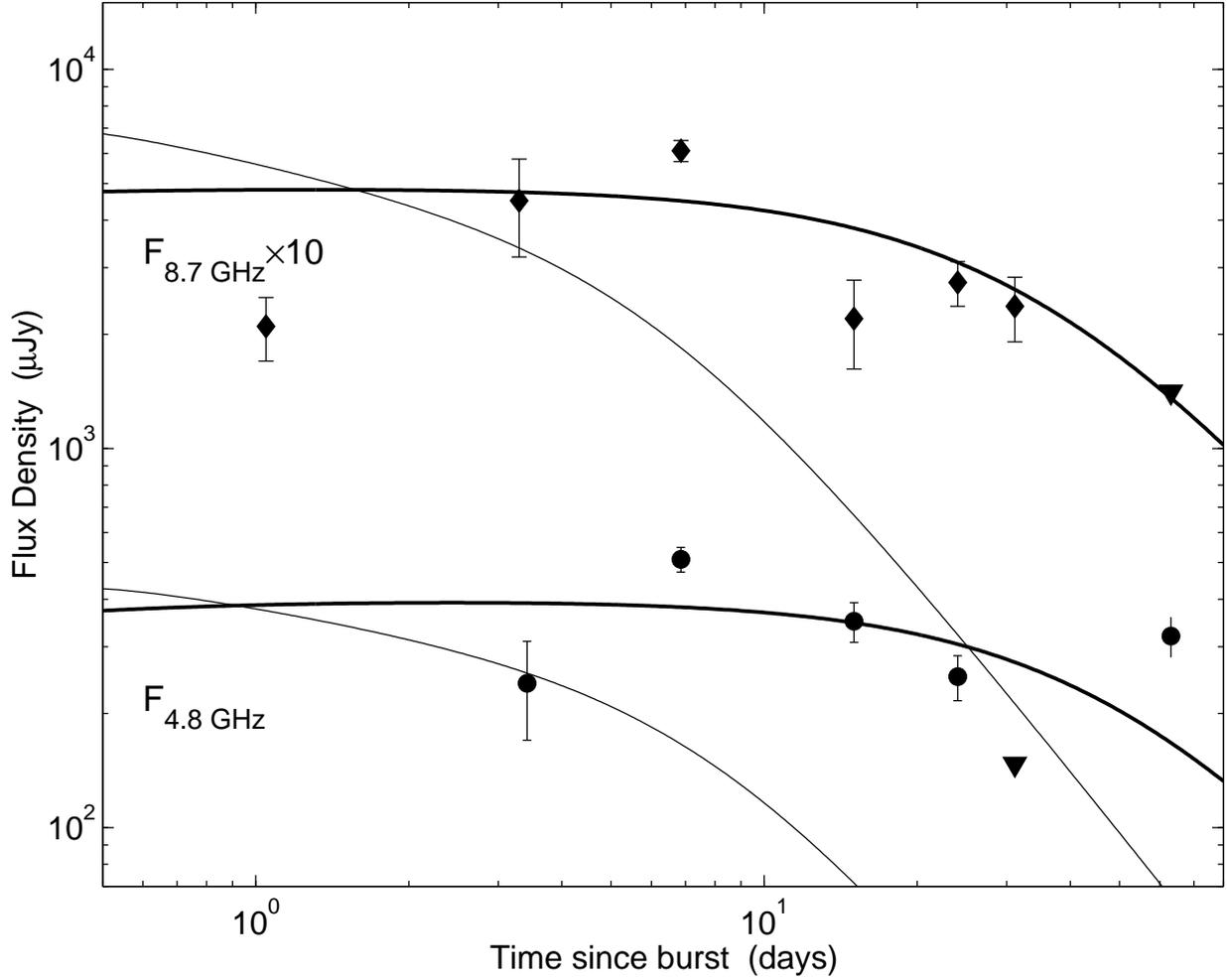}
\caption{The radio light curve of the afterglow of GRB~011121.  The
solid line is our wind model, the thin line is a representative jet
model, which is clearly excluded by the data.  The radio data exhibit
strong modulation due to interstellar scintillation and, as a result,
deviate from our model by more than $1\sigma$.  This will be addressed
in a future paper.}
\label{fig:radio}
\end{figure}

The radio measurements, however, do not show any sign of a decay until
at least $\sim 7$ days after the burst (Figure~\ref{fig:radio}).  The
rising centimeter-band flux prior to this time indicates that the jet
break is not at early times, and hence model {\it B}, the jet model,
is ruled out. This then leaves us with model {\it A}, the wind model.

\subsection{A Wind Model}

The multi-wavelength data, radio through optical, can only be analyzed
by considering the evolution of the broad-band synchrotron spectrum.
In addition to $\nu_c$, $p$ and $A_V$ we need to consider the
self-absorption frequency, $\nu_a$, and the peak frequency, $\nu_m$,
as well as the peak flux, $F_{\nu,m}$.  These parameters are estimated
from the data and can be inverted to yield physical quantities,
i.e.~the energy of the fireball, the density of the ambient medium,
and the fractions of energy in the electrons, $\epsilon_e$, and
magnetic field, $\epsilon_B$.  An example of this approach can be
found in \citet{bdf+01}.  The density in the wind model is
parameterized by $A_*$, which is defined through $A={\dot{M}}/{4\pi
v_w}=5\times 10^{11}A_*$ g/cm (see \citealt{cl99}) where $v_w$ is the
wind speed and $\dot M$ is the mass loss rate. The normalization of
$A_*=1$ applies for a typical Wolf-Rayet wind speed of $10^3$ km/sec
and $\dot{M}=10^{-5}$ M$_\odot$/yr.

Given the sparse data we prefer to undertake the model fitting in an
evolutionary approach rather than performing a blind $\chi^2$
minimization search.  For example, we fix the value of $p$ and $A_V$
to that determined earlier since the radio data has little bearing on
these parameters.  Next, we know that $\nu_c>\nu_O$, but there are no
data in the X-ray band to actually constrain the value of $\nu_c$.  We
therefore use $\nu_c\approx 10^{15}$ Hz since this is effectively the
lowest value the cooling frequency can have in this model. We will, at
a later point, revisit this issue and examine the consequences of
increasing $\nu_c$.

The remaining free parameters\footnotemark\footnotetext{Unless
otherwise stated, all time-dependent parameters are evaluated at epoch
1 day e.g.~$\nu_m\equiv\nu_m(t=1\,{\rm d})$.}, $\nu_a$, $\nu_m$, and
$F_{\nu,m}$ are relatively easy to constrain for the following
reasons.  The value of $F_{\nu,m}$ determines the overall scaling in
both the optical/NIR and radio bands, and is therefore constrained by
two sets of data.  The value of $\nu_m$ is constrained by the turnover
in the radio lightcurves (at $t\approx 7$ days; see
Figure~\ref{fig:radio}), as well as the flux density of the
optical/NIR lightcurves, since for a given value of $F_{\nu,m}$, the
flux density in the optical/NIR band is determined by $\nu_m$.

Finally, $\nu_a$ is constrained by the spectral slope between the two
centimeter bands.  The comparable flux between 4.8 and 8.7 GHz
suggests that $\nu_a < 4.8 GHz$.  An independent constraint on $\nu_a$
is also provided by the equation due to \citet{se01}:
\begin{equation} C=0.06(1+z)^4t_{\rm day}^4d_{\rm L,28}^{-2}\eta
\left(\frac{\nu_a}{{\rm GHz}}\right)^{10/3}
\left(\frac{\nu_m}{10^{13}\,{\rm Hz}}\right)^{13/6}
\left(\frac{\nu_c}{10^{14}\,{\rm Hz}}\right)^{3/2}
\left(\frac{F_{\nu,m}}{{\rm mJy}}\right)^{-1}\leq 0.25. 
\label{eqn:c}
\end{equation}
where $\eta = {\rm min}[(\nu_c/\nu_m)^{-(p-2)/2},1]$ is the fraction
of the electron energy radiated away.

We find $F_{\nu,m}\approx 3$ mJy, $\nu_c \approx 10^{15}$ Hz,
$\nu_m\approx 2\times 10^{12}$ Hz, $\nu_a\approx 1.4$ GHz and $\eta =
0.2$ provide an adequate description of the afterglow data.  From
these parameters we obtain $A_* \sim 0.01$ and that inverse Compton
cooling is marginally important.  Higher values of $A_*$ are possible
in the inverse Compton-dominated regime and if $C << 1$, implying that
$\nu_a$ is well below the centimeter bands.  With these observations,
we are unable to constrain such a model.


\section{Discussion \&\ Conclusions}
\label{sec:discussion}

GRB 011121, a relatively nearby burst ($z=0.36$), has shot to fame
given what appears to be firm identification of an underlying
supernova component \citep{bkp+02}.  Here we presented early time NIR
and comprehensive dual-frequency cm-wave observations of the
afterglow.  Thanks to the NIR data, we have been able to accurately
measure the considerable Galactic extinction towards the burst,
$A_V=1.16\pm 0.25$ mag, significantly smaller than that derived from
extrapolations of the IR maps \citep{sfd98}.  Our value of $A_V$ is an
important physical parameter in the modeling of the underlying SN
component \citep{bkp+02}.

If indeed long duration gamma-ray events such as GRB~011121 are linked
to SNe then the progenitors of GRBs are massive stars.  Such stars
possess strong winds and one expects to see a signature of the
wind-fed circumburst medium \citep{cl99}.  The optical/NIR data alone
rule out an isotropic explosion in a constant circumburst medium
model.  The radio data firmly rule out a model in which a jet is
fully-developed at $t<0.5$ d, but allow for a wind-fed circumburst
medium.  We estimate the mass loss rate, $\dot M \ale 10^{-7}
v_{w3}^{-1}\, M_\odot {\rm yr}^{-1}$ where $v_{w3}$ is the wind speed
in units of $10^3\,$km s$^{-1}$.  In the collapsar model
\citep{mwh01}, one expects the progenitors of GRBs to be massive stars
which have lost their hydrogen envelopes, i.e.~Wolf-Rayet stars.  For
such stars, $v_w\sim 10^3\,$km s$^{-1}$.

\begin{figure}[tbp]
\begin{center}
\includegraphics[angle=-90,scale=0.75]{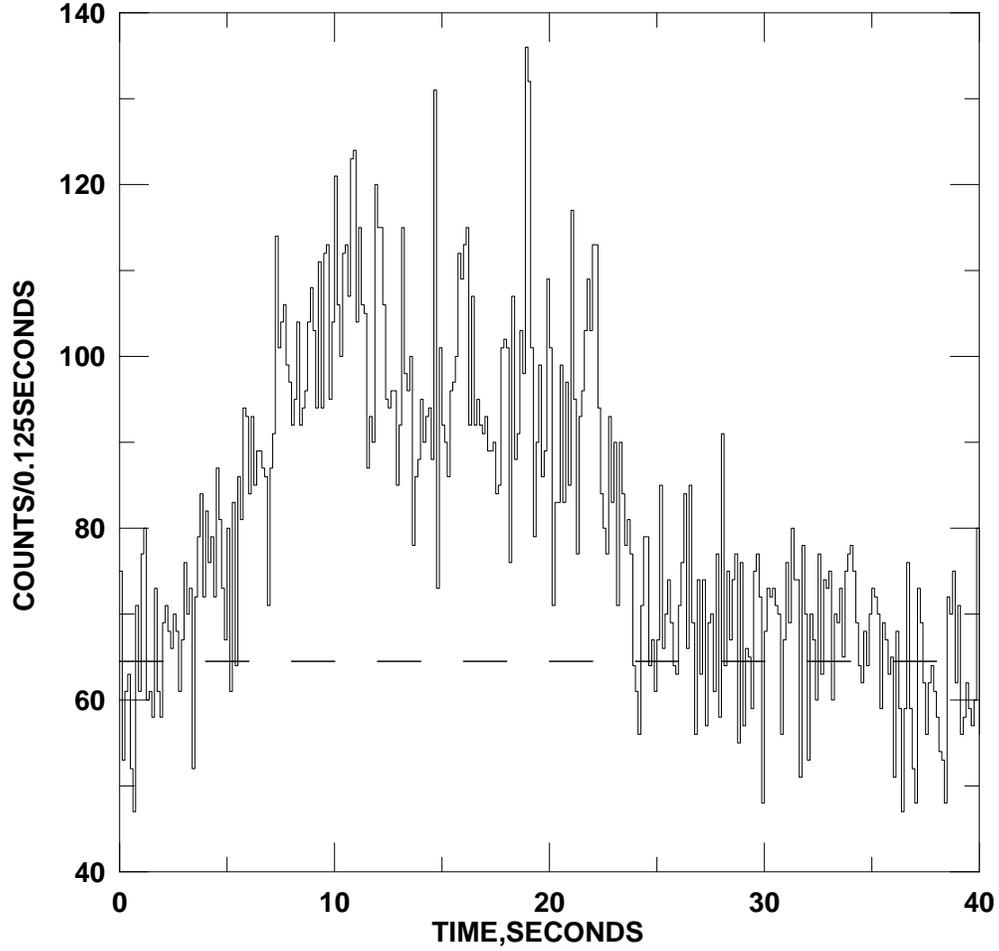}
\end{center}
\caption{Time history of GRB~011121 in the 25-150 keV energy range, as
observed by Ulysses.  The dashed line gives the background rate.  Zero
on the time axis corresponds to an Earth-crossing time of 67630.899 s.
}
\label{fig:grb-lc}
\end{figure}

Interestingly enough, this mass loss rate is similar to that inferred
for the progenitor of the Type Ic SN~1998bw, $2.5\times
10^{-7}$M$_\odot$\,yr$^{-1}$ \citep{cl99b}, based on the analysis of
the radio light curves \citep{kfw+98}.  This unusual SN is thought to
be associated with GRB~980425 based on spatial and temporal
coincidence \citep{gvv+98}, as well as its relativistic outflows
\citep{kfw+98}.  However, this GRB, if associated with SN~1998bw (as
we believe), releases at least three orders of magnitude less energy
in gamma-rays compared to cosmological bursts \citep{gvv+98} such as
GRB~011121. So the relation of GRB~980425 to cosmologically located
GRBs is unclear.  Nonetheless, we make the following curious
observation: the $\gamma$-ray profile (Figure~\ref{fig:grb-lc}) is of
similar duration and smoothness (with a few spikes superposed) as that
of GRB~980425.

The current data clearly rule out a jet break on the timescale of the
optical data, $t_j\age 2$ d, and the radio data require $t_j\age 7$
d. In the formulation of \citet{fks+01} the opening angle of the jet
must be wider than $\theta_j> 10$ degrees and hence the true energy
release is larger than $5 \times 10^{50}$ erg.  This lower limit is
consistent with the the clustering of energies around $5\times
10^{50}$ erg found by \citet{fks+01}.

Further improvements to the modeling is possible by including the
BeppoSAX measurement of the X-ray afterglow \citep{psa01}; the X-ray
flux will pin down $\nu_c$ quite well. We also note that the radio
fluxes given in Table~\ref{tab:radio} suffer from strong variability
(due to interstellar scintillation). Here we have used the mean
fluxes, and in a later paper we intend to report detailed analysis of
the scintillation and include the variability as a part of our
afterglow modeling, in particular as a way to constrain the size of
the afterglow region (c.f. \citealt{fkn+97}).

Thus, at least for one long duration burst the SN-GRB connection and a
massive progenitor origin appears to to have been established.
However, the true story may be more complex.  The absence of SN
components in other GRBs can be explained by appealing to the well
known wide diversity in luminosity of Type Ib/c SNe.  However, some of
the intensively observed afterglows are best modeled by expansion into
a homogeneous medium.  There could well be two different classes of
progenitors within the class of long-duration GRBs \citep{cl00}.

\acknowledgements

PAP gratefully acknowledges an Alex Rodgers Travelling Scholarship.
GRB research at Caltech (SRK, SGD, FAH, RS) is supported by grants
from NSF and NASA.  JSB is a Fannie and Hertz Foundation Fellow.  RS
holds a holds a Senior Fairchild Fellowship.  KH is grateful for
Ulysses support under JPL Contract 958056, and for IPN support under
NASA Grants FDNAG 5-11451 and NAG~5-10710.  We thank R.~Chevalier for
useful discussion.  Finally, we thank the staff of Las Campanas
Observatory and the ATNF for their assistance, and applaud the heroic
efforts of the staff of the AAT in obtaining these observations during
the commissioning of IRIS2.

\bibliographystyle{apj1b}
\bibliography{journals_apj,refs}

\clearpage

\begin{deluxetable}{lcrc}
\footnotesize
\tablecolumns{5}
\tablewidth{0pt}
\tablecaption{NIR observations of the afterglow of GRB~011121.}
\tablehead{\colhead{Date (2001 UT)} & \colhead{Filter} & \colhead{Magnitude} & \colhead{Telescope}}
\startdata
Nov 22.3560  &  $J$  &      17.852 $\pm$ 0.045  &	dP   \\
Nov 22.3573  &  $J$  &      17.730 $\pm$ 0.037  &	dP   \\
Nov 22.3587  &  $J$  &      17.763 $\pm$ 0.044  &	dP   \\
Nov 22.3600  &  $J$  &      17.801 $\pm$ 0.040  &	dP   \\
Nov 22.3614  &  $J$  &      17.821 $\pm$ 0.037  &	dP   \\
Nov 22.3627  &  $J$  &      17.799 $\pm$ 0.039  &	dP   \\
Nov 22.3641  &  $J$  &      17.785 $\pm$ 0.035  &	dP   \\
Nov 22.3654  &  $J$  &      17.770 $\pm$ 0.036  &	dP   \\
Nov 22.3667  &  $J$  &      17.795 $\pm$ 0.041  &	dP   \\
Nov 22.3681  &  $J$  &      17.739 $\pm$ 0.038  &	dP   \\
Nov 22.7177  &  $J$  &      18.352 $\pm$ 0.100  &	AAT   \\
Nov 23.3193  &  $J$  &      19.463 $\pm$ 0.068  &	dP   \\
Nov 28.5     &	$J$  &	    21.291 $\pm$ 0.282  &	Baade   \\
Nov 22.3178  &	$K$  &      15.959 $\pm$ 0.045  &	dP   \\
Nov 22.3194  &	$K$  &      15.987 $\pm$ 0.040  &	dP   \\
Nov 22.3211  &	$K$  &      15.908 $\pm$ 0.037  &	dP   \\
Nov 22.3227  &	$K$  &      15.994 $\pm$ 0.040  &	dP   \\
Nov 22.3244  &	$K$  &      15.958 $\pm$ 0.040  &	dP   \\
Nov 22.3263  &	$K$  &      16.002 $\pm$ 0.041  &	dP   \\
Nov 22.3279  &	$K$  &      16.006 $\pm$ 0.041  &	dP   \\
Nov 22.3296  &	$K$  &      16.003 $\pm$ 0.039  &	dP   \\
Nov 22.3296  &	$K$  &      16.003 $\pm$ 0.039  &	dP   \\
Nov 22.3313  &	$K$  &      15.981 $\pm$ 0.037  &	dP   \\
Nov 22.3329  &	$K$  &      16.053 $\pm$ 0.039  &	dP   \\
Nov 22.3349  &	$K$  &      16.039 $\pm$ 0.040  &	dP   \\
Nov 22.3365  &	$K$  &      15.997 $\pm$ 0.039  &	dP   \\
Nov 22.3382  &	$K$  &      16.120 $\pm$ 0.041  &	dP   \\
Nov 22.3398  &	$K$  &      15.996 $\pm$ 0.063  &	dP   \\
Nov 22.3454  &	$K$  &      16.027 $\pm$ 0.038  &	dP   \\
Nov 22.3470  &	$K$  &      16.069 $\pm$ 0.036  &	dP   \\
Nov 22.3487  &	$K$  &      16.100 $\pm$ 0.042  &	dP   \\
Nov 22.3503  &	$K$  &      16.015 $\pm$ 0.043  &	dP   \\
Nov 22.3520  &	$K$  &      16.098 $\pm$ 0.043  &	dP   \\
Nov 22.4771  &	$K$  &      16.421 $\pm$ 0.041  &	AAT   \\
Nov 22.4954  &	$K$  &      16.537 $\pm$ 0.041  &	AAT   \\
Nov 22.5126  &	$K$  &      16.495 $\pm$ 0.035  &	AAT   \\
Nov 22.6066  &	$K$  &      16.605 $\pm$ 0.058  &	AAT   \\
Nov 22.6169  &	$K$  &      16.788 $\pm$ 0.042  &	AAT   \\
Nov 22.6397  &	$K$  &      16.782 $\pm$ 0.038  &	AAT   \\
Nov 22.6506  &	$K$  &      16.862 $\pm$ 0.036  &	AAT   \\
Nov 22.6612  &	$K$  &      17.019 $\pm$ 0.052  &	AAT   \\
Nov 22.6716  &	$K$  &      16.852 $\pm$ 0.039  &	AAT   \\
Nov 22.6822  &	$K$  &      17.035 $\pm$ 0.083  &	AAT   \\
Nov 22.7272  &	$K$  &      17.005 $\pm$ 0.051  &	AAT   \\
Nov 22.7384  &	$K$  &      17.087 $\pm$ 0.079  &	AAT   \\
Nov 23.3336  &	$K$  &      17.924 $\pm$ 0.051  &	dP   \\
Nov 28.7092  &	$K$  &      19.346 $\pm$ 0.234  &	AAT   \\
\enddata
\tablecomments{(a)
Observations at the du~Pont (dP) 2.5-m were made by K.\ Koviak;
observations at the AAT were made by S.D.\ Ryder (Nov 22) and K.\ Gunn
(Nov 28); observations at the Baade telescope were made by M. Phillips.
(b) The following reference stars were used. For $K_s$ observations
on the AAT we observed UKIRT Faint Standards FS 7, 11 and 13 \citep{hll+01} on 2001 Nov.~28. 
SJ9113 \citep{pmkr+98} was observed at the du~Pont
telescope on 2001 Nov.~23.  We assumed an atmospheric extinction
coefficient of 0.09 mag/airmass in $K$ for the du~Pont observations,
and that the colour terms were negligible. 
We used the reference stars calibrated by \citet{pkgh+01} to
calibrate our $J$-band observations. 
(c) The AAT measurement of Nov. 28th is contaminated both by the host
and the nearby star.
}
\label{tab:nir}
\end{deluxetable}

\clearpage

\begin{deluxetable}{lcr}
\tabcolsep0in
\footnotesize
\tablewidth{0pt}
\tablecaption{Radio Observations of GRB~011121 made with the Australia
Telescope Compact Array.\label{tab:radio}}
\tablehead {
\colhead {Epoch}      &
\colhead {$\nu_0$} &
\colhead {S$\pm\sigma$} \\
\colhead {(UT)}      &
\colhead {(GHz)} &
\colhead {($\mu$Jy)}
}
\startdata
2001 Nov 25.20  &  4.80  &  240  $\pm$   70   \\
2001 Nov 28.64  &  4.80  &  510  $\pm$   38   \\
2001 Dec  6.80  &  4.80  &  350  $\pm$   42   \\
2001 Dec 15.80  &  4.80  &  250  $\pm$   34   \\
2001 Dec 22.90  &  4.80  &  -99  $\pm$   49   \\
2002 Jan 23.85  &  4.80  &  320  $\pm$   38   \\
2001 Nov 22.83  &  8.70  &  210  $\pm$   40   \\
2001 Nov 25.08  &  8.70  &  450  $\pm$  130   \\
2001 Nov 28.64  &  8.70  &  610  $\pm$   39   \\
2001 Dec  6.80  &  8.70  &  220  $\pm$   58   \\
2001 Dec 15.80  &  8.70  &  274  $\pm$   37   \\
2001 Dec 22.90  &  8.70  &  237  $\pm$   46   \\
2002 Jan 23.85  &  8.70  &  -99  $\pm$   47   \\
\enddata
\tablecomments{The columns are (left to right), UT date of the
start of each observation, center frequency, and peak flux density
at the best fit position of the radio transient, with the error given as
the root mean square noise on the image.
All observations were obtained using the continuum
mode and a 128 MHz bandwidth.  Flux calibration was
performed using PKS B1934$-$638, while the phase
was monitored using PKS B1057$-$797.
}
\end{deluxetable}

\clearpage

\begin{deluxetable}{cccrr}
\footnotesize
\tablecolumns{6}
\tablewidth{0pt}
\tablecaption{\label{tab:p}Afterglow Model Testing}
\tablehead{\colhead{Model} & \colhead{$\nu_c$} & \colhead{$(b,c)$} 
& \colhead{Closure} & \colhead{$p$} }
\startdata
ISM	&B & $(-3/2,0)$	   & $1.04 \pm 0.47$  & $3.21 \pm 0.08$   \\
ISM	&R & $(-3/2,1/2)$  & $2.04 \pm 0.47$  & $2.88 \pm 0.08$   \\
Wind	&B & $(-3/2,-1/2)$ & $0.04 \pm 0.47$  & $2.55 \pm 0.08$   \\
Wind	&R & $(-3/2,1/2)$  & $2.04 \pm 0.47$  & $2.88 \pm 0.08$ \\
Jet	&B & $(-2,-1)$	   & $-0.86 \pm 0.31$ & $1.66 \pm 0.06$   \\
Jet	&R & $(-2,0)$	   & $0.14 \pm 0.31$  & $1.66 \pm 0.06$ \\
\enddata
\tablecomments{Calculation of the closure relations $\alpha + b\beta +
c$ for a variety of afterglow models.  A successful model will have a
value of zero for the closure relation.  The ISM and Wind models are
for isotropic expansion in an homogeneous and wind-stratified medium
respectively.  The Jet model is for collimated expansion, with the jet
break time before the first observations were made.  The relations are
dependent on the assumed location of the cooling frequency, $\nu_c$
relative to the optical and NIR bands, $\nu_O$: the case $\nu_c>\nu_O$
is denoted by ``B''(for blueward) and $\nu_c<\nu_O$ by ``R'' (for
redward). $p$ is the electron energy power law index.
}
\end{deluxetable}

\clearpage

\clearpage

%

\clearpage

\end{document}